\documentclass[twocolumn,prl,aps,longbibliography,superscriptaddress,floatfix,tightenlines]{revtex4-2}
\usepackage{amsmath}
\usepackage{amssymb}
\usepackage{mathrsfs}
\usepackage{amsfonts}
\usepackage{epsfig}
\usepackage{bm}
\usepackage{times}
\usepackage{lineno}
\usepackage[T1]{fontenc}
\usepackage[dvips]{color}
\usepackage{braket}
\usepackage{setspace}
\usepackage[colorlinks,bookmarks=false,citecolor=blue,linkcolor=blue,urlcolor=blue]{hyperref}
\definecolor{darkblue}{rgb}{0.0, 0.0, 0.6}
\usepackage{titlesec}

\begin{document}
		\title{Global Precision Limits in Critical Quantum Metrology: From Cram\'{e}r-Rao to Ziv-Zakai}
	\author{Neng Zeng}
	\affiliation{School of Physics and Optoelectronics, South China University of Technology, Guangzhou 510640, China}
	
	\author{Tao Liu}
	\affiliation{School of Physics and Optoelectronics, South China University of Technology, Guangzhou 510640, China}
	
	\author{Yu-Ran Zhang}
	\email{yuranzhang@scut.edu.cn}
	\affiliation{School of Physics and Optoelectronics, South China University of Technology, Guangzhou 510640, China}
	\date{\today}
\begin{abstract}
Critical quantum metrology with equilibrium states predicts quantum-enhanced sensitivity only in the vicinity of criticality, where large prior information about the parameter is required.
By employing quantum Ziv-Zakai bounds, we derive a limit on the mean-square error in critical quantum metrology.
For second-order quantum phase transitions, we show that the precision predicted by the Cram\'{e}r-Rao bound offers no substantial improvement over the prior standard deviation.
Thus, the critical quantum sensor's precision can only achieve a constant gain compared to the prior standard deviation, even without performing any measurement.
We elucidate the fundamental limitation on the achievable precision in critical quantum metrology in the context of local sensing, even without considering state-preparation costs or noise.
Thus, the super-Heisenberg-limited sensitivity at criticality arises from precise prior knowledge rather than a genuine gain due to criticality.
Our work provides a practical framework for assessing critical quantum metrology and a routine for studying quantum sensing with many-body systems.
\end{abstract}

\maketitle 

 \emph{Introduction.---}%
Quantum metrology, harnessing quantum resources to enhance the sensitivity and precision of measurements \cite{Giovannetti2006,Giovannetti2011,Degen2017}, has been applied in various areas,
including optical interferometry \cite{Degen2017,Hosten2008,Haine2016,Helm2018}, atomic magnetometry \cite{Savukov2005,Budker2007,Aigner2008,Kitching2011}, gravitational-wave detection \cite{Abbott2009,Abadie2011,Aasi2013,Adhikari2014}, and atomic clocks \cite{Ludlow2015,Derevianko2011,Zaporski2025}.
The quantum Cram\'{e}r-Rao bound (QCRB) \cite{Braunstein1994,Ma2011}, with the quantum Fisher information (QFI), sets a limit on the standard deviation (SD) of the unbiased estimates in quantum parameter estimation,  $\delta \theta \ge \sqrt{\nu F_Q}$, with $\nu$ being the repetition number of measurements.
The exploitation of quantum resources, such as entanglement \cite{Giovannetti2004,Mitchell2004,Ono2013,Ma2011}
and quantum coherence \cite{Streltsov2017}, allows for the measurement precision beyond the standard quantum limit (SQL), $\delta \theta \propto N^{-1/2}$, and approaching the Heisenberg limit (HL), $\delta \theta \propto N^{-1}$, with $N$ being the number of elementary subsystems  constituting the probe.

Recently, criticality of quantum many-body systems has been proposed as a resource  in quantum parameter estimation for beating the SQL \cite{Ding2022,Frerot2018,Invernizzi2008,Zanardi2008,Garbe2020,Chu2021,Rams2018,Montenegro2021,Mukhopadhyay2024,Sarkar2024,Hotter2024,Liu2021,DiFresco2022,Ilias2022} and achieving the super-HL \cite{Xu2025,Yousefjani2023,He2023,Sahoo2025,Agarwal2025,Cheng2025}.
We focus on the approach of accessing the relevant physical parameter through a direct measurement of the many-body ground state \cite{Xu2025,Yousefjani2023,Liu2021,He2023,Sahoo2025,Agarwal2025,Cheng2025,DiFresco2022,Ilias2022}.
When a second-order quantum phase transition occurs in a quantum many-body system, the QFI for the ground state diverges near the critical point~\cite{CamposVenuti2007,Schwandt2009,Albuquerque2010,Gu2008,Greschner2013}, which implies an arbitrarily high measurement precision.
Unlike the interferometry-based scheme~\cite{Frerot2018}, critical quantum metrology offers a distinct paradigm capable of exhibiting super-HL~\cite{Yousefjani2023,He2023,Sahoo2025}, thereby attracting significant attention.

However, whether criticality-based sensing can, in practice, beat the HL remains a subject of debate.
First, resources for preparing large-scale entangled states, especially the time overhead, should be taken into consideration~\cite{Chu2023}.
Due to the vanishing energy gap at criticality~\cite{Zurek2005,Dziarmaga2010}, adiabatic state preparation would be extremely time-consuming~\cite{Rams2018,Gietka2021}, revealing a fundamental constraint on the feasibility of critical quantum metrology. 
For many-body probes, since the substantial resource costs limit the number of measurement trials, the QCRB grossly overestimates the achievable precision, despite being asymptotically tight
\cite{Trees2001,Trees2007,Tsang2012,Hai-Long2026}.
Second, noise could also be amplified by divergent phenomena at criticality, compromising the sensitivity benefit of critical quantum metrology
\cite{Wiersig2020,Loughlin2024,Zheng2025,Mihailescu2025,He2023a,Mihailescu2026,Chen2021}.
Moreover, regardless of the state preparation and noise, the saturation of the QCRB for critical quantum metrology requires the assumption of the unbiased estimation, which usually cannot be fulfilled in experiments~ \cite{Ziv1969}.
Critical quantum metrology is only beneficial in the context of the \textit{local} estimation that requires large prior information about the parameter, as the effective parameter regime for criticality-enhanced sensitivity shrinks sharply with the system size.
Thus, as a \textit{global} bound on mean-square error (MSE), that rigorously accounts for prior information and remains valid for biased estimators, the quantum Ziv-Zakai bound (QZZB) 
\cite{Trees2007,Ziv1969,Chazan1975,Bell1997,Tsang2012,Giovannetti2012}
provides a tighter and more practical limit than QCRB, thereby emerging as a powerful tool for investigating critical quantum metrology.

In this Letter, by applying the QZZB to account for prior information about the parameter, we obtain a lower bound on the MSE in critical quantum metrology.
To quantify the practical gain of critical quantum metrology relative to the prior information, we introduce an enhancement factor $\eta$, defined as the ratio of the prior SD to the sensor's precision, and derive an upper bound on it. 
We show that, for the ground state during the second-order quantum phase transitions, the enhancement factor near criticality is bounded by a constant independent of the system size $N$ and the spatial dimension $d$.
For concreteness, we consider a one-dimensional (1D) system with Stark localization and an anisotropic transverse-field XY chain, for which the QCRB predicts ultrahigh precision near the critical points, reaching or even beating the HL.
However, when the prior information required for the super-HL sensitivity is taken into consideration, the net gains in estimation precision for these two models remain constant compared to their prior SD. 
Therefore, the criticality-enhancement is not a genuine gain from criticality, but a consequence of the precise prior knowledge.
Our work provides a useful framework for assessing the practical potential of critical quantum sensors, guiding future investigations of quantum sensing with many-body systems.

 \emph{Quantum Cram\'{e}r-Rao bound and critical quantum metrology.---}%
Quantum parameter estimation aims to estimate an unknown parameter $\theta$ imprinted on a quantum state $\rho_\theta$, with a set of POVMs $\{\hat{E}(x)\}$ yielding outcomes ${x}$ with probability $P(x|\theta) = \textrm{tr}[\rho_\theta \hat{E}(x)]$.
For an unbiased estimator $\hat{\theta}$, the variance $(\delta \theta)^2\equiv{\langle \hat{\theta}^2 \rangle-\langle \hat{\theta}\rangle^2}$ is lower bounded by the CRB, i.e., $(\delta \theta)^2\ge1/{F[\rho_\theta]}$, with $F[\rho_\theta]\equiv\int \! dx \;  [\partial_\theta P(x|\theta)]^2/P(x|\theta)$
being the Fisher information (FI)~\cite{Helstrom1969,Wiseman2009}.
The QFI refers to the maximum FI over all possible POVMs~\cite{Braunstein1994}, i.e.,
$F_Q(\theta) \equiv \max_{\{\hat{E}(x)\}} F[\rho_\theta]$.
For a pure state $\rho_\theta=\ket{\psi_\theta}\bra{\psi_\theta}$, the QFI is simplified as $F_Q[\ket{\psi_\theta}]=4(\langle  \partial_\theta\psi_\theta|\partial_\theta\psi_\theta\rangle+|\langle\psi_\theta|\partial_\theta\psi_\theta\rangle|^2)$~\cite{Giovannetti2006,Giovannetti2011}.

We consider a system of a size $N$, described by a Hamiltonian $H_\theta$, where the target parameter $\theta$ drives a quantum phase translation at the critical point $\theta_c$. 
Within the critical quantum metrology framework, measurements are performed on the non-degenerate ground state $\ket{\psi_\theta}$.
The QFI for $\ket{\psi_\theta}$ relates to fidelity susceptibility as $F_Q(\theta)=
8\lim_{\delta\theta\to0} (1-\mathcal{F}(\rho_\theta,\rho_{\theta+\delta \theta}))/\delta\theta^2$, under the condition of $\textrm{Tr}(\partial_\theta \rho_\theta)=0$, with the fidelity $\mathcal{F}(\rho_\theta,\rho_{\theta+\delta \theta})=\textrm{Tr}({\sqrt{\rho_\theta}\rho_{\theta+\delta \theta}\sqrt{\rho_\theta}})^{1/2}$~\cite{You2007}.
Therefore, the divergent fidelity susceptibility of the ground state in the vicinity of criticality leads to the divergence of the QFI, implying ultrahigh estimation sensitivity.
In the thermodynamic limit $N\to\infty$, the QFI scales as $F_Q(\theta)\propto |\theta-\theta_c|^{-\alpha}$, with $\alpha$ being the critical exponent~\cite{CamposVenuti2007,Schwandt2009,Gu2008,Albuquerque2010,Greschner2013}.
In a finite-size system, consisting of $N=L^d$ subsystems with $d$ being the spatial dimension, the QFI at the critical point scales as $F_Q(\theta_c,N)\propto N^{\beta/d}$.
Furthermore, these two asymptotic behaviors can be merged into an ansatz
$F_Q(\theta,N)\propto(N^{-\beta/d}+C|\theta-\theta_c|^\alpha)^{-1}$~\cite{He2023}.
In addition, for second-order quantum phase transitions, the scale invariance dictates a conventional finite-size scaling ansatz for the QFI, i.e, $F_Q(\theta)=N^{\alpha/\nu}f(N^{1/\nu}(\theta-\theta_c))$, where $f(x)$ is a function depending on the model and vanishes for $x \to \infty$ \cite{Pathria2011}.
Note that the critical exponents $\beta$, $\alpha$, and $\nu$ are constrained by a formula $\beta/d=\alpha/\nu$ \cite{Rams2018,Schwandt2009,Albuquerque2010}.

To evaluate the critical parameter region for the criticality-enhanced sensitivity, we define the critical width $W_c(N)$ considering $|\theta - \theta_c| < W_c(N)$, under which the QFI scales as 
\begin{equation}
	F_Q(\theta,N)\propto N^{\beta/d}= N^{\alpha/(\nu d)}. \label{1}
\end{equation}
From this, one readily finds $W_c(N) \propto N^{-1/(\nu d)}$.
Therefore, when $|\theta - \theta_c| < W_c$, a system with a large $\alpha/(\nu d)$ can lead to an ultra-sensitive estimation of $\theta$. 
Furthermore, when $\alpha/(\nu d) > 2$, it implies the capability of achieving the super-HL~\cite{Yousefjani2023,He2023,Sahoo2025,Agarwal2025,Cheng2025}.
However, this scaling is confined to a critical width $W_c$ around $\theta_c$ that shrinks with the increase of the system size $N$, thereby demanding increasingly precise prior information.
It is therefore significant to study whether the criticality-enhanced estimation precision is an artifact of the extremely high degree of prior information about the parameter of interest.
This problem, about prior information about the parameter, is beyond the local estimation theory characterized by the QCRB and inspires the consideration of the QZZB that is accessible for global estimation.

 \emph{Quantum Ziv-Zakai Bound.---}%
The Ziv-Zakai Bound (ZZB) provides a lower bound on the MSE by connecting parameter estimation to the hypothesis testing problem~\cite{Ziv1969}.
Its quantum counterpart~\cite{Tsang2012} yields a global and non-asymptotic bound that inherently incorporates prior information and remains valid for biased estimators.
Compared to the QCRB, the QZZB provides a more suitable framework for assessing the practical performance of critical quantum metrology.

The QZZB evaluates the estimation precision by the MSE,
	$\Sigma=\int_{-\infty}^{+\infty} \! d\theta P(\theta)[\int_{-\infty}^{+\infty} \! d\boldsymbol{x} P(\boldsymbol{x}|\theta)[\hat{\theta}(\boldsymbol{x})-\theta]^2]$,
where $\hat{\theta}$ is the estimator, and $P(\theta)$ denotes prior distribution.
The MSE is lower bounded by the QZZB as~\cite{Trees2007,Tsang2012}:
\begin{equation}
	\begin{split}
		\Sigma\ge \Sigma_z=\frac{1}{2}\int^{+\infty}_0 \!\!\!\!\! d\tau \;\!&\tau \int^{+\infty}_{-\infty} \!\!\!\! d\theta \min (P_\theta(\theta),P_\theta(\theta+\tau))\\ &\quad\times[1-\sqrt{1-\mathcal{F}^2(\rho_\theta,\rho_{\theta+\tau})}].            \label{2}
	\end{split}
\end{equation}
Consider a uniform prior distribution as $P_\theta(\theta)=\frac{1}{W}\textrm{rect}(\frac{\theta-\theta_c}{W})$, with a mean $\mu=\theta_c$ and width $W$. See the Supplemental Material (SM)~\cite{SM} for the discussions of the Gaussian prior distribution.
For the system ground state $\ket{\psi_\theta}$, the QZZB can be explicitly rewritten as
\begin{equation}
	\begin{split}
		\Sigma_z= \frac{1}{2W}\int^{W}_0 \!\!\! d\tau \!\;\tau \!\int^{\theta_c+W/2-\tau}_{\theta_c-W/2} \!\!\!\! d\theta\; (1-\sqrt{1-|\langle \psi_\theta|\psi_{\theta+\tau}\rangle|^2}).     \label{3}      
	\end{split}
\end{equation}

In the high prior information (HPI) regime, corresponding to a small $W$, $\Sigma_z$ is completely dominated by the prior variance, i.e., $\lim_{W\to0}\Sigma_z=W^2/12$ \cite{Giovannetti2012}. 
One can obtain a comparable precision at criticality as predicted by the QCRB, $1/F_Q(\theta_c)$, just using HPI without performing any measurement.
Moreover, in the low prior information (LPI) regime, non-critical regions far from criticality dominate the QZZB, $\Sigma_z$, so that the critical contribution becomes negligible.
Thus, to realize criticality-enhancement, the prior width $W$ should be comparable to $W_c(N)$, avoiding both the dilution from non-critical regions and the trivial precision gain from an overly narrow prior.

Since critical quantum metrology only offers an advantage within a narrow regime sufficiently for local estimation, it is essential to quantify the precision enhancement provided by the QCRB compared to the prior SD.
First, we employ the QZZB to find the prior width $W(N)$ that enables $\Sigma_z(N,W(N))=1/F_Q(\theta_c,N)$ for a fixed $N$.
We define an enhancement factor $\eta$, which quantifies the nut gain in the measurement precision, $\delta\theta_{\textrm{min}}=1/\sqrt{F_Q}$, over the prior SD, $\delta\theta_{\textrm{prior}}$, as 
\begin{equation}
	\eta \equiv \frac{\delta\theta_{\textrm{prior}}}{\delta \theta_{\textrm{min}}} = W\sqrt{\frac{F_Q}{12}}.
\end{equation}
Then, the criticality-enhancement can be  achieved if the prior width $W$ satisfies the local estimation condition $W<W_u$, where the upper bound can be expressed as $W_u(N)=AN^{-1/(\nu d)}$, in terms of a constant $A$, is proportional to the critical width $W_c(N)$.
Thus, with Eq.~(\ref{1}), we obtain an upper bound on $\eta$: 
\begin{equation}
	\eta\le\frac{A}{\sqrt{12B}}N^{(\alpha/2-1)/ (\nu d)}, \label{7}
\end{equation}
where $F_Q(\theta_c,N)=B^{-1}N^{\alpha/(\nu d)}$, with $B$ being a constant.
Moreover, for a system exhibiting second-order quantum phase transitions in its ground state, it holds universally that $\alpha = 2$~\cite{Schwandt2009,Albuquerque2010}.
In this case, we obtain that
\begin{equation}
	\eta \le \frac{A}{\sqrt{12B}}, \label{6}
\end{equation}
showing that the enhancement factor $\eta$ is upper bounded by a constant $\eta_{\textrm{max}}\equiv A/\sqrt{12B}$.

This result shows that for second-order phase transitions, critical quantum metrology at most provides a constant improvement over the prior SD, independent of the system size $N$ and the spatial dimension $d$.
Moreover, our framework generalizes to noisy scenarios~\cite{Zhang2014}, which even degrades the achievable precision, thereby reducing the upper bound on the enhancement factor to a lower constant.
Next, we investigate two models that are proposed to achieve or even beat the HL, including a 1D system with Stark localization~\cite{He2023} and the anisotropic transverse-field XY chain~\cite{Mukhopadhyay2024}.

\begin{figure} [h]
	\centering
	\includegraphics[width=0.48\textwidth]{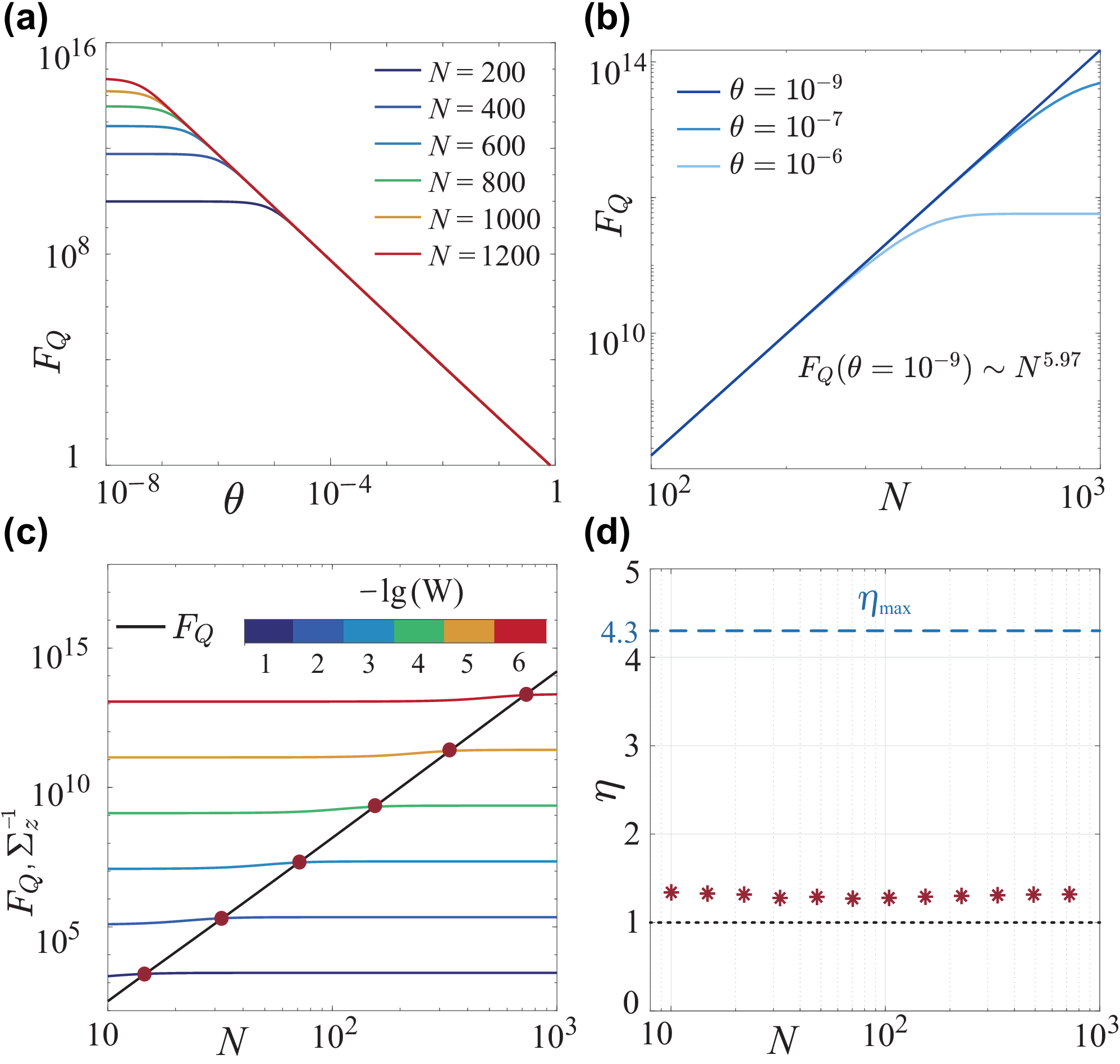}\\
	\caption
	{ 1D system with Stark localization (\ref{4}).
	(a) QFI versus $\theta$ for the ground state with different system sizes $N$.
	(b) QFI versus the system size $N$ for the ground state for $\theta=10^{-9}$, $10^{-7}$, and $10^{-6}$. 
	The data are well fitted by $F_Q \propto N^{\beta}$, where the scaling exponent $\beta=5.97$.
	(c) Inverse QZZB $\Sigma_z^{-1}$ versus $N$ for different prior widths $W$, compared to the QFI (black solid curve).
	(d) Enhancement factor $\eta$ (red scatters) versus the system size $N$, evaluated at the points when the QZZB coincides with the QCRB [red dots in (c)], i.e., $\Sigma_z=F_Q^{-1}$. 
	The enhancement factor $\eta$ is around $1.3$, that is independent of the system size $N$. 
	It demonstrates that the improvement in estimation precision over the prior uncertainty is limited to a constant factor.
	The blue horizontal dashed line represents the derived upper bound $\eta_{\textrm{max}} \equiv A/\sqrt{12B} \approx 4.3$.}
	\label{fig:1}
\end{figure}

 \emph{1D system with Stark localization.---}%
We consider a chain of $N$ sites with a tunneling rate $J=1$, under a gradient field $\theta$ to be estimated~\cite{He2023}, of which the Hamiltonian is
\begin{equation}
	H(\theta)=\sum_{i=1}^{N-1}\ket{i}\bra{i+1}+\ket{i+1}\bra{i}+\theta\sum_{i=1}^{N}i\ket{i}\bra{i}. \label{4}
\end{equation}
The system undergoes a phase transition from an extended phase to a Stark localized phase, which in the thermodynamic limit ($N\to\infty$) occurs at $\theta=\theta_c=0$~\cite{Kolovsky2008,Schulz2019}.
The QFI peaks at the critical point $\theta_c$ and follows a finite-size scaling ansatz $F_Q(\theta, N) \propto [N^{-\beta/(\nu d)} + C|\theta - \theta_c|^\alpha ]^{-1}$.
Inside the critical parameter region $|\theta - \theta_c| < W_c$, the QFI scales as $F_Q \propto N^{\beta}$, whereas outside this region it follows $F_Q \propto |\theta - \theta_c|^{-\alpha}$, see Fig.~\ref{fig:1}(a) and \ref{fig:1}(b).
By employing the ansatz $F_Q(\theta) = N^{\alpha/\nu} f(N^{1/\nu}(\theta-\theta_c))$, the critical point and the critical exponents are determined as $(\theta_c, \alpha, \nu) = (10^{-9}, 2.00, 0.335)$~\cite{He2023}.
Then, using the QZZB for a uniform prior distribution (\ref{3}), we calculate a lower bound on the MSE for the model. 
Due to the Hamiltonian's symmetry, we restrict our analysis to the domain $\theta>0$.
Given that the critical point is located at $\theta_c=10^{-9}$, the prior distribution is adjusted to $P_\theta=(1/W)\textrm{rect}[(\theta-W/2)/W]$, with $W \ge 10^{-8}$ ensuring coverage of $\theta_c$.
Figure~\ref{fig:1}(c) shows the inverse QZZB $\Sigma_z^{-1}$ as a function of $N$ for different prior widths $W$, revealing that the QZZB is governed by $W$, with only a weak dependence on $N$.
A faint increase in $\Sigma_z^{-1}$ with $N$ emerges only within a narrow range of $N$, where the critical length $W_c$ matches the scale of $W$.

We next investigate the enhancement factor $\eta$ where the QZZB coincides with the QCRB, i.e., $\Sigma_z(N,W) =1/ F_Q(\theta_c,N)$.
We need to determine the values of $A$ and $B$ before evaluating the upper bound $\eta_{\textrm{max}}$.
By analyzing the scaling of QFI at the critical point, $F_Q(\theta_c, N) = B^{-1} N^{\alpha / (\nu d)}$, as shown in Fig.~\ref{fig:1}(b), we obtain the value $B \approx 4167$.
To determine $A$, we need to identify the upper bound $W_u$, within which the criticality-enhancement occurs.
We introduce an indicator function $G(N,W)=-\delta\lg\Sigma_z(N,W)/\delta\lg N$, which increases with the decrease of $W$ and quantifies the degree of criticality-enhancement (see SM~\cite{SM} for more details).
For a fixed $N$, $G(N,W)$ converges to a non-critical baseline $\tilde{G}(N)$ as $W\to\infty$.
We define $W_u$ as the threshold, where the criticality-enhancement vanishes, and set without loss of generality that $G(N, W_u) = 0.1 \max_{W} G(N, W) + 0.9\tilde{G}(N)$.
The extracted $W_u$ exhibits a scaling behavior $W_u = A N^{-1/\nu}$, with $A \approx 963.4$.
This leads to an upper bound of $\eta_{\textrm{max}} \approx 4.3$.
To validate Eq.~(\ref{6}), we compute the enhancement factor $\eta$ at each intersection point [$\Sigma_z=F_Q^{-1}$, red dots in Fig~\ref{fig:1}(c)]. 
All obtained values are around $1.3$ below the upper bound $4.3$, exhibiting only a constant enhancement in precision of critical quantum metrology over the prior SD, see Fig.~\ref{fig:1}(d).
This result demonstrates that although the QCRB suggests a super-HL, to achieve this in practice requires that the prior information distribution is already comparable to the super-HL precision.
Hence, the net gain over the prior SD is only a constant factor of $1.3$ in this case, which agrees well with our theoretical predictions.

\begin{figure}
	\centering
	\includegraphics[width=0.48\textwidth]{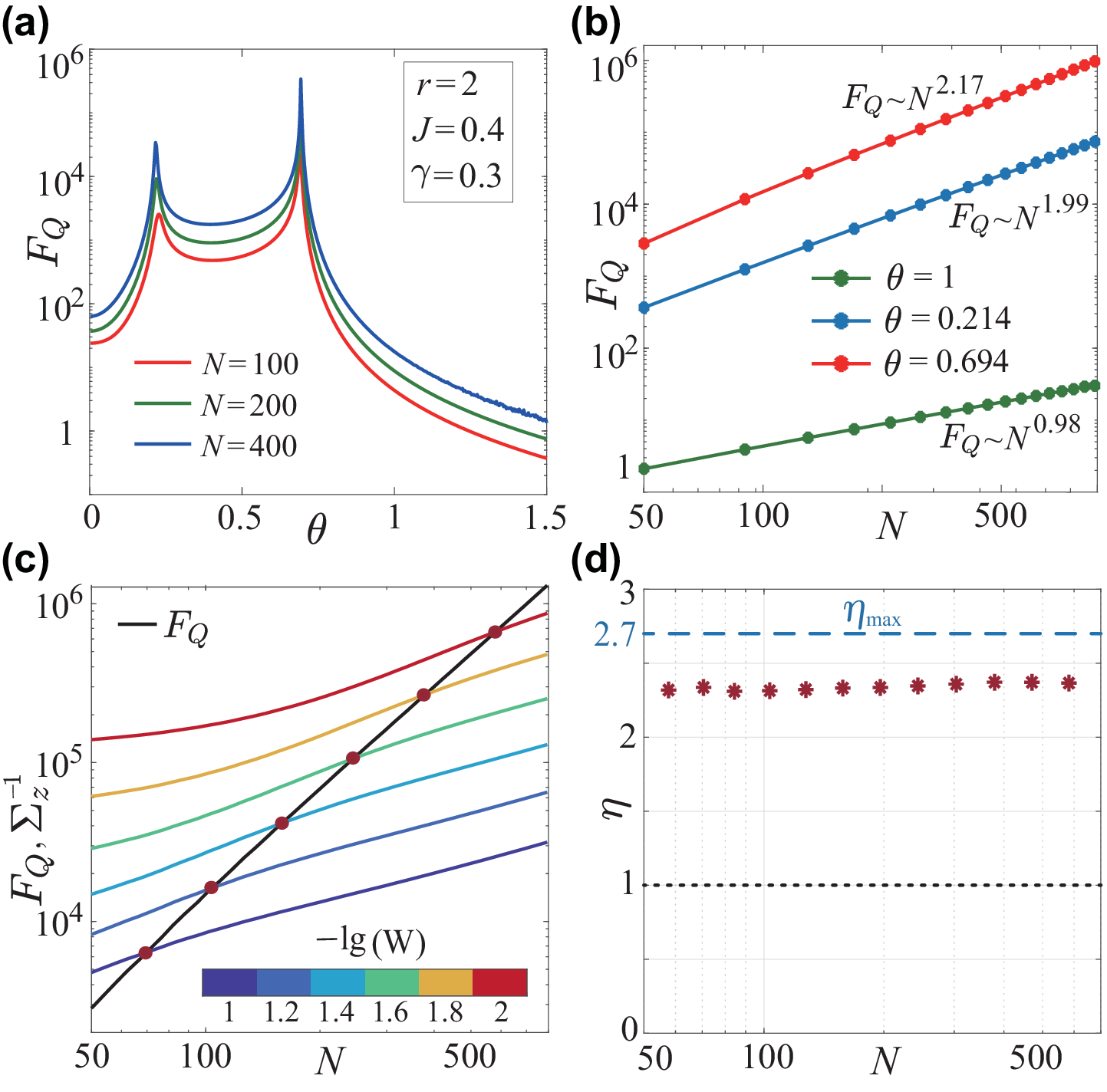}\\
	\caption
	{  Anisotropic transverse-field XY chain (\ref{5}).
		(a) QFI versus $\theta$ for different system sizes $N$.
		 QFI peaks at both critical points, $\theta=0.214$ and $0.694$.
		(b) QFI versus the system size $N$ at $\theta=1$, $0.214$, and $0.694$, and the scaling of the QFI follows $F_Q \sim N^{0.98}$, $N^{1.99}$, and $N^{2.17}$, respectively.
		(c) Inverse QZZB, $\Sigma_z^{-1}$, versus $N$ for different prior widths $W$, compared to the QFI (black solid curve).
		(d) Enhancement factor $\eta$ (red scatters) versus the system size $N$, evaluated at the intersection points [red dots in (c)], when the QZZB coincides with the QCRB, i.e., $\Sigma_z=F_Q^{-1}$. 
		The enhancement factor $\eta$ is around $2.4$ and independent of the system size $N$. 
		It confirms that attaining the QCRB precision leads to only a constant enhancement compared to the prior SD.
		The blue horizontal dashed line represents the upper bound $\eta_{\textrm{max}} \equiv A/\sqrt{12B} \approx 2.7$. 
	}
	\label{fig:2}
\end{figure}

 \emph{Anisotropic transverse-field XY chain.---}%
We next investigate an anisotropic transverse-field XY chain~\cite{Mukhopadhyay2024}, of which the Hamiltonian is given by
\begin{equation}
	H=-\sum_i \frac{J_i}{2} [(1+\gamma)\sigma_i^x\sigma_{i+1}^x+(1-\gamma)\sigma_i^y\sigma_{i+1}^y]+\theta\sum_i \sigma_i^z, \label{5}
\end{equation}
where $J_i$ denotes the nearest-neighbor interaction strength, $\gamma$ is the anisotropic factor, and $\theta$ represents the external field to be estimated.
The system consists of multiple cells of size $r$, with the intracell coupling fixed at $J_0= 1$ and the intercell coupling being an adjustable parameter $J_i = J$.
By setting the parameters as $r=2$, $\gamma=0.3$, and $J=0.4$, the system undergoes phase transitions at $\theta=0.214$ and $\theta=0.694$, where the QFI for the ground state peaks at either critical point, see Fig.~\ref{fig:2}(a).
We focus on the scaling of the QFI with the increase of $N$ at two critical points and compare it to the case far from criticality ($\theta=1$). 
Away from the critical point, the scaling of the QFI follows the SQL. 
In comparisons, at the critical points, $\theta=0.214$ and $\theta=0.694$, it reaches the HL, see Fig.~\ref{fig:2}(b).

We next employ the QZZB for a uniform prior distribution (\ref{3}) on the MSE of critical quantum sensing with the anisotropic transverse-field XY chain.
We show $\Sigma_z$ against the system size $N$ for different prior widths $W$ at the critical point $\theta=0.694$ in Fig.~\ref{fig:2}(c).
In contrast to the HL scaling of sensitivity due to the QCRB, $\Sigma_z$ exhibits a sub-SQL scaling, following $\Sigma_z \propto N^{0.7}$.
Similarly, we obtain an upper bound for the enhancement factor $\eta$ as $\eta_{\textrm{max}} \approx 2.7$, where $A \approx 12.2$ and $B \approx 1.69$.
We evaluate $\eta$ at the intersection points $\Sigma_z=F_Q^{-1}$ [red dots in Fig~\ref{fig:2}(c)] for different prior widths $W$.
All values of $\eta$ lie around 2.4, remaining below the upper bound of $\eta_{\textrm{max}}=2.7$, see Fig.~\ref{fig:2}(d).

Thus, our results for both models show that when the prior distribution is taken into account, critical quantum metrology can only achieve a constant gain of the precision compared to the prior SD, which is the prior precision without any measurement.
Furthermore, since the QZZBs are generally not tight, the values of the enhancement factors may even be overstated, and the practical gain could be smaller.

\emph{Conclusion and discussion.---}%
In summary, by applying the QZZB that accounts for the prior information about the parameter of interest, we have established a fundamental limit on the precision for critical quantum metrology.
We consider the enhancement factor $\eta$, defined as the ratio of the estimation precision to the prior SD for the scenario, where the QZZB coincides with the QCRB.
We derive an upper bound $\eta_{\textrm{max}}$ of the enhancement factor, showing that for generic second-order ground-state phase transitions, i.e., $\alpha=2$, the upper bound $\eta_{\textrm{max}}$ is a constant independent of the system size $N$.
It shows that the critical quantum sensors with second-order phase transitions can only achieve a constant precision gain over the prior SD.
Although the QCRB would suggest super-HL scaling at criticality, its realization requires prior information that already provides the same super-HL precision without any measurement.
Consequently, the super-HL enhancement of sensitivity from criticality is attributable to the precise prior knowledge, not an intrinsic benefit of criticality.
Our work establishes a practical framework for assessing criticality-based quantum metrology, which also accounts for the effects of noise and would be helpful for the investigation of critical quantum metrology in the presence of noise.

Note that even though the QCRB tends to overestimate the performance of criticality-based sensors, this does not mean they are fundamentally ineffective. 
While our work establishes a general limit for prevalent second-order transitions, sensing schemes based on first-order \cite{Campostrini2014} or topological phase transitions \cite{Budich2020,Sarkar2022,Koch2022} constitute an open area for future research. 
Furthermore, multiparameter estimation at critical points remains a promising frontier, with recent work suggesting that criticality may mitigate inherent estimation incompatibilities \cite{DiFresco2022}. 
Finally, the characteristics of equilibrium criticality transcend static properties and persist in non-equilibrium regimes.
Therefore, critical quantum dynamics may offer promising schemes for critical quantum metrology with many-body systems~\cite{Mishra2021,Chu2021,Guan2021,Li2026}.

\begin{acknowledgments}
T.L. acknowledges the support from the National Natural Science Foundation of China (Grant No.~12274142), 
Guangdong Provincial Quantum Science Strategic Initiative (Grant No.~GDZX2505004), 
and Introduced Innovative Team Project of Guangdong Pearl River Talents Program (Grant No.~2021ZT09Z109).
Y.R.Z. is supported in part by:
the National Natural Science Foundation of China (Grant No.~12475017), 
Scientific Research Innovation Capability Support Project for Young Faculty (Grant No.~SRICSPYF-ZY2025171), 
the Natural Science Foundation of Guangdong Province (Grant No.~2024A1515010398), 
and the Startup Grant of South China University of Technology (Grant No.~20240061).
\end{acknowledgments}


\bibliography{QCM.bib}

\end{document}


\title{SUPPLEMENTAL MATERIAL:\\Global Precision Limits in Critical Quantum Metrology: From Cram\'{e}r-Rao to Ziv-Zakai}
	\author{Neng Zeng}
	\affiliation{School of Physics and Optoelectronics, South China University of Technology, Guangzhou 510640, China}
	
	\author{Tao Liu}
	\affiliation{School of Physics and Optoelectronics, South China University of Technology, Guangzhou 510640, China}
	
	\author{Yu-Ran Zhang}
	\email{yuranzhang@scut.edu.cn}
	\affiliation{School of Physics and Optoelectronics, South China University of Technology, Guangzhou 510640, China}
	
	\date{\today}
	\begin{abstract}
		In the Supplemental Material, we provide detailed derivations of the quantum Ziv-Zakai bound. 
		Then, we explain the indicator function $G$ introduced in the main text, which quantifies the degree of criticality-enhancement.
		Through the color maps of $G$, we show that criticality-enhancement vanishes in the high and low prior information regimes. 
		In addition, we consider the performance of the two models under a Gaussian prior distribution.
	\end{abstract}
	
	\maketitle
	

	\renewcommand{\theequation}{S\arabic{equation}}
	\setcounter{equation}{0}  
	
	\renewcommand{\thefigure}{S\arabic{figure}}
	\setcounter{figure}{0}  
	
	\renewcommand{\thetable}{S\arabic{table}}
	\setcounter{table}{0}
	
	
\section{Derivation of the Quantum Ziv-Zakai Bound}
	 
 Let $\boldsymbol{\xi}=\{\xi_1,\cdots,\xi_j\}$ denote the measurement outcomes, $\hat{\theta}(\boldsymbol{\xi})$ be the estimator, $\epsilon(\boldsymbol{\xi}) \equiv |\hat{\theta} (\boldsymbol{\xi})- \theta|$ be the estimation error, and $P_\theta (\theta)$ the prior distribution. 
 The mean squared error (MSE) is given by
 \begin{equation}
 	\begin{split}
	 \Sigma&\equiv \int_{-\infty}^{+\infty} \!\! \!\!\! dx \;\! d\boldsymbol{\xi}\; P_\theta(x) P_{\boldsymbol{\xi}|\theta}(\boldsymbol{\xi}|x) \epsilon(\boldsymbol{\xi})^2\\
	 &= \int_{-\infty}^{+\infty} \!\!\!\!\! dx \;\! P_\theta(x) \int_0^{+\infty} \!\!\!\! d\epsilon \;\! P_{\epsilon|\theta}(\epsilon|x) \epsilon^2. \label{1}
	\end{split} 
\end{equation}
The probability density of $\epsilon$ is the derivative of the cumulative distribution function, i.e., $P_\epsilon(s) = \frac{d}{ds} \text{Pr}(\epsilon \le s)=-\frac{d}{ds} \text{Pr}(\epsilon \ge s)$. Thus, we have
 \begin{equation}
	\begin{split}
		\int_0^{+\infty} \!\!\!\! d\epsilon\;\! P_{\epsilon|\theta}(\epsilon|\theta) \epsilon^2&=\int_0^{+\infty} \!\!\!\! ds \!\; 2s\text{Pr}(|\hat{\theta} - \theta|\ge s|\theta)\\
		&=\int_0^{+\infty} \!\!\!\! d\tau \!\; \frac{\tau}{2}\text{Pr}(|\hat{\theta} - \theta|\ge \frac{\tau}{2}|\theta), \label{2}
	\end{split}
\end{equation}	 
where $\tau=2s$.
\begin{table}[b]
	\centering
	\caption{Binary hypothesis test}
	\label{table}
	\setlength{\tabcolsep}{5pt}
	\renewcommand{\arraystretch}{1.2}
	\begin{tabular}{lcc}
		\hline
		\noalign{\vskip -2pt}
		\hline
		\noalign{\vskip 3pt}  
		\textbf{Two hypotheses} & \(\mathcal{H}^0: \theta = x\) & \(\mathcal{H}^1: \theta = x + \tau\) \\
		\midrule
		Probability & \(\text{Pr}(\mathcal{H}^0) = P_0\) & \(\text{Pr}(\mathcal{H}^1) = P_1\) \\
		Decision rule & \( \theta(\boldsymbol{\xi}) \leq x+ \frac{\tau}{2}\) & \(\theta(\boldsymbol{\xi}) \geq x + \frac{\tau}{2}\) \\
		\noalign{\vskip 3pt}
		\hline
		\noalign{\vskip -2pt}
		\hline 
	\end{tabular}
\end{table}
Substituting Eq.~(\ref{2}) into Eq.~(\ref{1}), a lower bound on the MSE can be expressed as an integral over the error probability of a binary hypothesis test~(\ref{table}) as
 \begin{equation}
	\begin{split}
		 \Sigma\!&=\!\!\int_0^{+\infty} \!\!\!\! d\tau \frac{\tau}{2} \int_{-\infty}^{+\infty} \!\! \! dx P_\theta(x) [\text{Pr}(\hat{\theta} \ge x\!+\!\frac{\tau}{2}|x)\!+\!\text{Pr}(\hat{\theta} \le x\!-\!\frac{\tau}{2}|x)]\\
		 &=\!\!\int_0^{+\infty} \!\!\! d\tau \frac{\tau}{2} \int_{-\infty}^{+\infty} \!\! \! dx~ [P_\theta(x)+P_\theta(x+\tau)]\\
		 &~~~~~~~~~~~~~~~~~~~~\times P_0 \text{Pr}(\hat{\theta} \ge x\!+\!\frac{\tau}{2}|x)+ P_1 \text{Pr}(\hat{\theta} \le x\!-\!\frac{\tau}{2}|x+\tau)\\
		 &=\!\!\int_0^{+\infty} \!\!\! d\tau \frac{\tau}{2} \int_{-\infty}^{+\infty} \!\! \! dx~ [P_\theta(x)+P_\theta(x+\tau)] \text{Pr}_e(x,x+\tau), \label{3}
	\end{split}
\end{equation}		 
with $P_0=\frac{P_\theta(x)}{P_\theta(x)+P_\theta(x+\tau)}$ and $P_1=\frac{P_\theta(x+\tau)}{P_\theta(x)+P_\theta(x+\tau)}$.
The error probability 
\begin{equation}
	\text{Pr}_e(x,x+\tau)=P_0 \text{Pr}(\hat{\theta} \ge x+\frac{\tau}{2}|x)+ P_1 \text{Pr}(\hat{\theta} \le x-\frac{\tau}{2}|x+\tau)
\end{equation}
 is lower bounded by the minimum error probability $\mathcal{P}_e(x,x+\tau)$, which is independent of the form of the estimator $\hat{\theta}(\boldsymbol{\xi})$.
Therefore, one obtains the Ziv-Zakai bound \cite{Trees2007} as
\begin{equation}
	\begin{split}
		\Sigma \ge \!\!\int_0^{+\infty} \!\!\! d\tau \frac{\tau}{2} \mathcal{V} \int_{-\infty}^{+\infty} \!\! \! dx [P_\theta(x)+P_\theta(x+\tau)] \mathcal{P}_e(x,x+\tau), \label{4}
	\end{split}
\end{equation}	
where $\mathcal{V}f(x)\equiv \max_{y\ge0} f(x+y)$ denotes the valley--filling operation.
An alternative form of the Ziv-Zakai bound is
\begin{equation}
	\begin{split}
		\Sigma \ge \!\!\int_0^{+\infty} \!\!\! d\tau \tau \mathcal{V} \int_{-\infty}^{+\infty} \!\! \! dx \min[P_\theta(x),P_\theta(x+\tau)] \mathcal{P}_e^{el}(x,x+\tau), \label{5}
	\end{split}
\end{equation}	
where $\mathcal{P}_e^{el}$ is the minimum error probability for equally likely hypotheses.

To obtain the lower bound on the error probability $\text{Pr}_e(x,x+\tau)$, we define a pair of projectors $\{\Pi_0,\Pi_1\}$ with $\Pi_0+\Pi_1=\mathbb{I}$, representing a two-outcome measurement.
Then, the error probability can be expressed as
\begin{equation}
	\begin{split}
		\text{Pr}_e(x,x+\tau)&=P_0\textrm{Tr}(\rho_x\Pi_1)+P_1\textrm{Tr}(\rho_{x+\tau}\Pi_0)\\
		&=P_0+\textrm{Tr}(\Gamma \Pi_0),
	\end{split}
\end{equation}
with $\Gamma=P_1\rho_{x+\tau}-P_0\rho_{x}$.
Therefore, the minimum error probability $\mathcal{P}_e(x,x+\tau)$ is obtained as
\begin{equation}
	  \mathcal{P}_e(x,x+\tau)\equiv\min_{\{\Pi_0,\Pi_1\}} P_0+\textrm{Tr}(\Gamma \Pi_0)= \frac{1}{2}-\frac{1}{2}\|\Gamma\|, \label{6}
\end{equation}	
where $\|\cdots\|$ denotes the trace norm.
The optimal measurement is given by $\Pi_0 = \sum_{{i|\gamma_i\le0}}\ket{i}\bra{i}$, where $\gamma_i$ are the eigenvalues of $\Gamma$ in its spectral decomposition $\Gamma=\sum_i \gamma_i \ket{i}\bra{i}$.
Furthermore, for equally likely hypotheses $P_0=P_1$, the minimum error probability can be further bounded by an expression involving the fidelity $\mathcal{F}$, given as
\begin{equation}
	\mathcal{P}_e^{el}(x,x+\tau)= \frac{1}{2}-\frac{1}{2}\|\Gamma\|\ge\frac{1}{2}(1-\sqrt{1-\mathcal{F}^2(\rho_x,\rho_{x+\tau})}), \label{7}
\end{equation}	
where $\mathcal{F}(\rho_x,\rho_{x+\tau})=\textrm{Tr}\sqrt{\sqrt{\rho_x}\rho_{x+\tau}\sqrt{\rho_x}}$.
Finally, we obtain a conventional form of the quantum Ziv-Zakai bound (QZZB) using Eq.~(\ref{4}) as~\cite{Tsang2012}
\begin{equation}
	\begin{split}
		\Sigma\ge \Sigma_z\equiv\frac{1}{2}\int^{+\infty}_0 \!\!\! d\tau &\tau  \mathcal{V} [\int^{+\infty}_{-\infty} \!\!\! dx \min (P_\theta(x),P_\theta(x+\tau))\\ &\times(1-\sqrt{1-\mathcal{F}^2(\rho_x,\rho_{x+\tau})})].            \label{8}
	\end{split}
\end{equation}
In practice, the valley-filling operation is often omitted to favor computational tractability at the expense of a looser bound. 
Another version of the QZZB, based on Eq.~(\ref{5}), is written as \cite{Giovannetti2012} 
\begin{equation}
	\begin{split}
			\Sigma\ge \frac{1}{4}\mathcal{V}\int^{+\infty}_0 \!\!\! d\tau &\tau  [\int^{+\infty}_{-\infty} \!\!\! dx (P_\theta(x)+P_\theta(x+\tau))\\ &\times(1-\sqrt{1-4P_0P_1\mathcal{F}^2(\rho_x,\rho_{x+\tau})})].            \label{9}
	\end{split}
\end{equation}
When the prior distribution $P_\theta$ is a uniform window, the two bounds in Eqs.~(\ref{8}) and~(\ref{9}), are equivalent.
Under a Gaussian prior distribution, the bound (\ref{9}) is tighter than the bound (\ref{8}).

\section{Regimes of high and low prior information}
In critical quantum metrology, the many-body ground state $\ket{\psi_\theta}$ is directly measured to estimate the unknown parameter $\theta$ imprinted in the system Hamiltonian $H_\theta$.
Here, we focus on the two models introduced in the main text.

\emph{Model 1---}%
The first model is a one-dimensional system with Stark localization of length $N$, described by the Hamiltonian 
\begin{equation}
	H(h)=\sum_{i=1}^{N-1}\ket{i}\bra{i+1}+\ket{i+1}\bra{i}+h\sum_{i=1}^{N}i\ket{i}\bra{i}. \label{10}
\end{equation} 
The gradient field $h$ serves as the parameter to be estimated, and the system exhibits a critical point at $h_c = 0$.

\emph{Model 2---}%
The second model is the anisotropic transverse-field XY chain of size $N$, described by the Hamiltonian
\begin{equation}
	H\!=\!-\frac{1}{2}\sum_i J_i ((1+\gamma)\sigma_i^x\sigma_{i+1}^x+(1-\gamma)\sigma_i^y\sigma_{i+1}^y)+h\sum_i \sigma_i^z, \label{11}
\end{equation}
where $J_i$ denotes the nearest-neighbor interaction strength, $\gamma$ is the anisotropic factor, and $h$ represents the external field to be estimated.
The system consists of multiple cells of size $r$, where the intracell coupling is fixed at $J_0=1$ and the intercell coupling is an adjustable parameter $J_i=J$. For the parameter set $r=2$, $\gamma=0.3$, and $J=0.4$, the system exhibits phase transitions at $h_c=0.214$ and $h_c=0.694$.
In the following, we focus on the critical point at $h_c=0.694$.

\begin{figure} [t]
	\centering
	\includegraphics[width=0.48\textwidth]{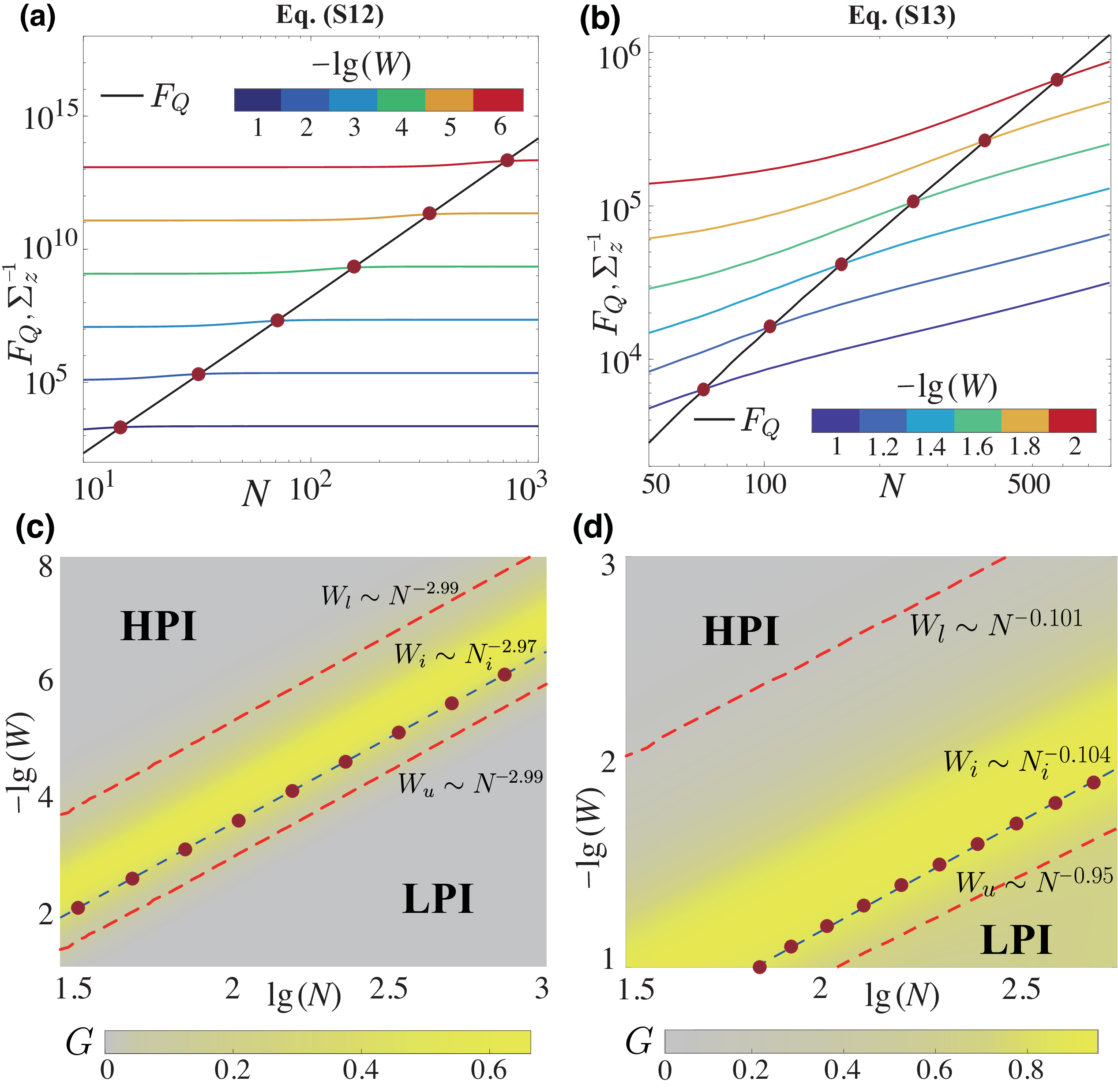}\\
	\caption
	{ Two models with Hamiltonians in Eqs.~(\ref{10}) and~(\ref{11}) for uniform prior distribution in Eq.~(\ref{14}). 
	(a) and (b) Quantum Fisher information (QFI) $F_Q$ and the inverse quantum Ziv-Zakai bound (QZZB) $\Sigma_z^{-1}$ for different prior widths $W$ in the models described by (\ref{10}) and (\ref{11}), respectively. 
	The intersection points $(W_i, N_i)$, satisfying $\Sigma_z=1/F_Q$ are marked by red dots.
	(c) and (d) Color maps of $G=-\delta\lg\Sigma_z/\delta\lg N$ in the $\lg N$--$\lg W$ plane for the models in Eqs.~(\ref{10}) and~(\ref{11}), respectively.
	The red dashed curves denote the upper bound $W_u$ and lower bound $W_l$.
	For both models, the bounds $W_l$ and $W_u$ approximately follow $W_u \sim W_l \sim N^{-1/(\nu d)}$, with $1/\nu \approx 2.99$ for the model in Eq.~(\ref{10}) and $1/\nu \approx 1$ for the model in Eq.~(\ref{11}).	The intersection points $(W_i, N_i)$ lie within the region $W_l < W < W_u$ and follow a scaling $W_i \sim N_i^{-1/(\nu d)}$.}
	\label{fig:2}
\end{figure}

First, we assume the prior distribution is a uniform window with mean $\mu=\theta_c$ and width $W$, i.e.,
\begin{equation}
 P_\theta(\theta)=\frac{1}{W}\textrm{rect}\left(\frac{\theta-\theta_c}{W}\right), \label{14}
\end{equation}
with $\theta_c$ being a critical point of the system.
Then, we apply the QZZB to the two models Eq.~(\ref{10}) and Eq.~(\ref{11}), and compute the lower bounds on the mean-square error $\Sigma_z(N,W)$ for different system sizes $N$ and prior widths $W$ [Fig.~\ref{fig:2}(a) and~\ref{fig:2}(b)].

To quantify the degree of criticality-enhancement, we consider the local scaling of the inverse QZZB, $\Sigma_z^{-1}(N,W)$, with the system size $N$, defined through the discrete derivative
\begin{equation}
     G(N,W) \equiv - \frac{\delta \lg \Sigma_z(N,W)}{\delta \lg N}.
\end{equation}
The behavior of $G$ depends strongly on the prior width $W$.
In the high prior information (HPI) regime where $W$ is small, $\Sigma_z$ is dominated by the prior variance and weakly depends on the system size $N$.
In the limit $W\to 0$, we have $\Sigma_z\to W^2/12$, so that $G\to 0$.
Moreover, in the low prior information (LPI) regime where $W$ is large,  the QZZB is dominated by non-critical regions.
In the extreme limit $W\to \infty$, $G(N,W)$ converges to a non-critical baseline $\tilde{G}(N)$, representing a purely size-dependent improvement of precision without criticality.
For a fixed $N$, when the prior width $W$ is comparable to the critical width $W_c(N)$, the system experiences the criticality-enhancement.
Here, $W_c(N)$ characterizes the critical region $|\theta - \theta_c| < W_c(N)$, in which the quantum Fisher information (QFI) scales as
\begin{equation}
	 F_Q(\theta,N)\propto N^{\beta/d}= N^{\alpha/(\nu d)}.
\end{equation}
In this case, $G(N,W)$ can significantly exceed the non-critical baseline $\tilde{G}(N)$.
Thus, the extent to what $G(N,W)$ rises above $\tilde{G}(N)$ serves as a quantification of the strength of criticality-enhancement, while in the HPI and LPI regimes, the critical contribution is negligible [see Fig.~\ref{fig:2}(c) and~\ref{fig:2}(d)].

Therefore, the criticality-enhancement in quantum metrology is contingent on a specific range of the prior width $W$ for a fixed system size $N$.
Moreover, for any fixed $N$, the prior width for criticality-enhancement satisfies $W_l < W < W_u$.
The lower bound $W_l$ is determined by the condition $W_l^2/12 = 1/F_Q$, ensuring that the prior variance does not surpass the Cram\'{e}r-Rao bound (QCRB) precision.
The upper bound $W_u$ is defined as the threshold, where the criticality-enhancement vanishes. 
Without loss of generality, we set
\begin{equation}
	G(N, W_u) = 0.1 \max_{W} G(N, W) + 0.9\tilde{G}(N).
\end{equation}
By extracting $W_u$ as a function of $N$ [see Fig.~\ref{fig:2}(c) and~\ref{fig:2}(d)], we find that for the two models in Eq.~(\ref{10}) and Eq.~(\ref{11}), $W_u$ exhibits the scaling behaviors $W_u\sim N^{-2.99}$ and $W_u\sim N^{-0.95}$, respectively.
To connect these observations to the critical properties of the system, we determine the critical exponent $\nu$ by employing the finite-size scaling ansatz 
\begin{equation}
	F_Q(\theta) = N^{\alpha/\nu} f(N^{1/\nu}(\theta-\theta_c)).
\end{equation}
For the two models, this yields $\nu=0.335$~\cite{He2023} and $\nu=1$~\cite{Mukhopadhyay2024}, respectively. 
Remarkably, both cases follow the same general scaling relation $W_u\sim N^{-1/(\nu d)}$, which is proportional to the critical width $W_u\sim N^{-1/(\nu d)}$, i.e., $W_u\sim W_c$.
Given that $F_Q\propto N^{\alpha/(\nu d)}$ in the vicinity of the critical points
\cite{Rams2018,Schwandt2009,Albuquerque2010},
the lower bound scales as 
\begin{equation}
	W_l \sim N^{-\alpha/(2\nu d)}.
\end{equation}
For second-order phase transitions ($\alpha=2$), we find 
\begin{equation}
	W_l \sim W_u \sim N^{-1/(\nu d)}.
\end{equation}

Then, we focus on the enhancement factor, defined as
\begin{equation}
         \eta\equiv\delta\theta_{\textrm{prior}}/\delta\theta_{\textrm{min}}=W\sqrt{F_Q/12}, 
\end{equation}
where $\delta\theta_{\textrm{prior}}$ is the prior standard deviation, and $\delta\theta_{\textrm{min}}=1/\sqrt{F_Q}$ is the QCRB precision. 
It quantifies the precision gain over the prior standard deviation when the QZZB coincides with the QCRB, i.e., $\Sigma_z=1/F_Q$.
We compute $F_Q(N)$ at the critical point and $\Sigma_z^{-1}(N)$ for different prior widths $W$ [see Fig.~\ref{fig:2}(a) and~\ref{fig:2}(b)]. 
The enhancement factor $\eta$ is then evaluated at each intersection point $(W_i, N_i)$, satisfying $\Sigma_z(N_i,W_i)=F_Q(N_i)$.
In the $\lg N$--$\lg W$ plane, all intersection points lie within the region $W_l < W < W_u$ and follow a scaling $W_i \sim N_i^{-1/(\nu d)}$, with $d=1$ [see Fig.~\ref{fig:2}(c) and~\ref{fig:2}(d)].
Moreover, substituting $W_i \sim N_i^{-1/(\nu d)}$ and $F_Q\sim N^{\alpha/(\nu d)}$ into the definition $\eta=W\sqrt{F_Q/12}$, we find
\begin{equation}
	\eta(W_i, N_i) = W_i\sqrt{\frac{F_Q(N_i)}{12}}\sim N_i^{(\alpha/2-1)/(\nu d)}. \label{12}
\end{equation}
For a second-order quantum phase transition ($\alpha=2$), the exponent vanishes, which is fully consistent with the conclusion that $\eta$ remains constant.

\begin{figure}[h]
	\centering
	\includegraphics[width=0.48\textwidth]{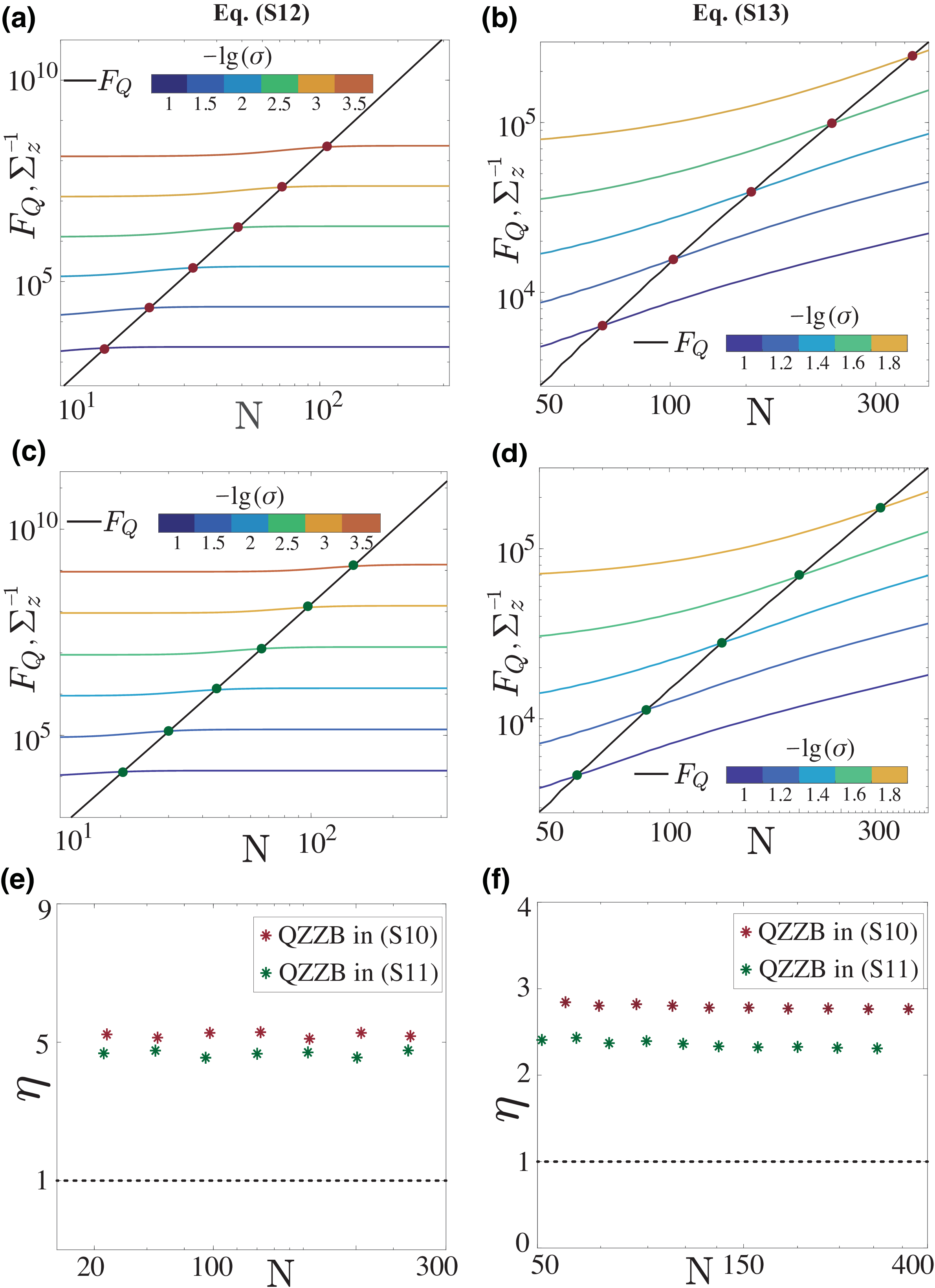}\\
	\caption
	{ 	Two models with Hamiltonians Eqs.~(\ref{10}) and~(\ref{11}) for Gaussian prior distribution in Eq.~(\ref{23}).
		(a) and (b) Inverse QZZB, $\Sigma_z^{-1}$, as a function of $N$ for different values of $\sigma$, computed with $F_Q$, using the equally-likely-hypothesis QZZB in Eq.~(\ref{8}) for the two models, respectively.
		(c) and (d) Same as (a) and (b), but using the tighter QZZB in Eq.~(\ref{9}).
		(e) and (f) Enhancement factor $\eta$ [red stars for the QZZB in Eq.~(\ref{8}) and green stars for that in Eq.~(\ref{9})] versus the system size $N$, evaluated at the intersection points, where the QZZB coincides with the QCRB [red dots in (a) and (b), green dots in (c) and (d)], i.e., $\Sigma_z=F_Q^{-1}$.
		For the model in Eq.~(\ref{10}), $\eta$ is around $5.2$ for the QZZB in Eq.~(\ref{8}) and $4.8$ for that in Eq.~(\ref{9}).
		For the model in Eq.~(\ref{11}), the corresponding values are around $2.8$ and $2.4$.
		The enhancement factors remain constant for each model under each QZZB version, and those obtained from the tighter QZZB in Eq.~(\ref{9}) are consistently smaller than those from Eq.~(\ref{8}).}
	\label{fig:3}
\end{figure}
\section{Analysis under Gaussian Prior Information }	  
Next, consider a Gaussian prior distribution $P_\theta(\theta)$ with its mean positioned at the critical point $\mu = \theta_c$ and variance $\sigma^2$, given by
\begin{equation}
	P_\theta(\theta) = \frac{1}{\sqrt{2\pi\sigma^2}} \exp\left[-\frac{(\theta - \theta_c)^2}{2\sigma^2}\right].  \label{23}
\end{equation}
Substituting this Gaussian prior into the QZZBs in Eqs.~(\ref{8}) and~(\ref{9}), we evaluate the lower bound $\Sigma_z$ on the mean-square error.
For simplicity, we omit the valley-filling operation $\mathcal{V}$.
Following the previous analysis, we compute the inverse QZZB $\Sigma_z^{-1}$ as a function of $N$ for different values of $\sigma$, for the two models in Eqs.~(\ref{10}) and~(\ref{11}), using the two QZZBs in Eqs.~(\ref{8}) and~(\ref{9}).
The results are shown in Fig.~\ref{fig:3}(a)-(d), computed to $F_Q$: (a) and (b) correspond to the QZZB in Eq.~(\ref{8}) for the two models, while (c) and (d) correspond to the QZZB in Eq.~(\ref{9}), which is tighter than Eq.~(\ref{8}) under the Gaussian prior.
The enhancement factor $\eta$ is obtained from the intersection points $(\sigma_i, N_i)$, when the QZZB coincides with the QCRB, i.e., $\Sigma_z=F_Q^{-1}$.
For a Gaussian prior, the prior variance is $(\delta\theta_{\textrm{prior}})^2=\sigma^2/8$, so the enhancement factor is
\begin{equation} 
	\eta\equiv\delta\theta_{\textrm{prior}}/\delta\theta_{\textrm{min}}=\sigma \sqrt{\frac{F_Q}{8}}.
\end{equation}
Following the same procedure, we compute $\eta$ from the intersection points using the two QZZBs for the two models~[see Fig.~\ref{fig:3}(e) and~\ref{fig:3}(f)].
As the uniform prior case, the enhancement factors remain constant for both models under two different versions of the QZZB. 
Moreover, for both models, the values of $\eta$ obtained from the tighter QZZB in Eq.~(\ref{9}) are smaller than those from Eq.~(\ref{8}).

\bibliography{QCM.bib}